\documentclass[aps,prl,10pt,twocolumn,superscriptaddress,longbibliography]{revtex4-1} 
\pdfoutput=1
\usepackage[dvips]{epsfig}
\usepackage{graphicx}  
\usepackage{dcolumn}   
\usepackage{bm}        
\usepackage{amssymb}   
\usepackage{upgreek}
\usepackage{color}
\usepackage{float}
\usepackage[linkcolor=blue,colorlinks=true,urlcolor=blue,citecolor=red]{hyperref}
\usepackage{braket}
\usepackage{subcaption}
\usepackage{amsmath}

\usepackage[nameinlink]{cleveref}

\begin{document}

\title{Slow propagation of 2 GHz acoustical waves in a suspended GaAs on insulator phononic waveguide}

\author{Giuseppe Modica}
\affiliation{Centre de Nanosciences et de Nanotechnologies, CNRS,  Universit\'{e} Paris-Saclay, Palaiseau, France}

\author{Rui Zhu}
\affiliation{Centre de Nanosciences et de Nanotechnologies, CNRS,  Universit\'{e} Paris-Saclay, Palaiseau, France}

\author{Robert Horvath}
\affiliation{Centre de Nanosciences et de Nanotechnologies, CNRS,  Universit\'{e} Paris-Saclay, Palaiseau, France}

\author{Gregoire Beaudoin}
\affiliation{Centre de Nanosciences et de Nanotechnologies, CNRS,  Universit\'{e} Paris-Saclay, Palaiseau, France}

\author{Isabelle Sagnes}
\affiliation{Centre de Nanosciences et de Nanotechnologies, CNRS,  Universit\'{e} Paris-Saclay, Palaiseau, France}

\author{R\'{e}my Braive}
\affiliation{Centre de Nanosciences et de Nanotechnologies, CNRS,  Universit\'{e} Paris-Saclay, Palaiseau, France}
\affiliation{Universit\'{e} de Paris, 75207 Paris Cedex 13, France}

\begin{abstract}
Optoelectronic oscillators have dominated the scene of microwave oscillators in the last few years thanks to their great performances regarding frequency stability and phase noise. However, miniaturization of such a device is an up to date challenge. Recently, devices based on phonon-photon interaction gather a lot of interest thanks to their extreme compactness and working frequency directly in the GHz. In this frame, a still missing element to obtain long-term frequency stability performances is an on-chip delay within the feedback loop. Here, we experimentally show filtering and slow propagation of 2 GHz acoustic waves on a Gallium Arsenide membrane heterogeneously integrated on silicon wafer. By engineering the dispersion of an acoustical waveguide, we evidence a group velocity below 1000 m/s for the mode able to propagate. Thus, an integrated delay implementation is at reach for potential improvement of opto-acoustic devices such as optomechanical oscillators or wireless applications.

\end{abstract}

\maketitle

Oscillators play a crucial role in modern society. Periodic signals represent basic ingredients for telecommunications and navigation systems \cite{rubiola2009phase} as well as sensing and metrology applications \cite{fan2017high, li2012femtometer, xu2017dual, lee2016displacement}.  Working frequency directly in the ultra high frequency regime (UHF) and high spectral purity are the essential features that must be addressed to strictly fulfill oscillator’s requirements in the current scenario. Optoelectronic oscillators (OEOs) have been a good answer to these needs, allowing better phase noise performances with respect to traditional microwave oscillators thanks to ultra-low loss fiber as energy storage element \cite{liu2018stable}. However, kilometers of fiber are usually needed to introduce the proper time delay to reach high quality factor for good phase noise performances.
This bulky elements expose the OEO to environmental perturbations such as thermal \cite{kaba2006improving,eliyahu2002improving} and mechanically induced fluctuations \cite{hati2008cancellation}. Therefore, compactness is a clue to face these limitations. One may look towards optomechanics which imprint mechanical frequencies on a optical carrier \cite{aspelmeyer2014cavity}. Optomechanical devices such as 1D optomechanical crystal have gained a lot of importance in the last few years thanks to their working frequency directly in the GHz regime \cite{eichenfield2009optomechanical}, compactness and their integrability \cite{tsvirkun2015integrated}. Thus, under certain conditions, optomechanical crystals fuelled by optical injection power, sustain oscillations directly in the GHz \cite{ghorbel2019optomechanical} also called optomechanical oscillators (OMOs).
With this technique the amount of power that can be injected is limited by thermo-optical instabilities, and the lack of a feedback loop prevents frequency and phase noise to be furthermore optimized in the long-term stability.
To push these limits beyond, another way than optics is required to control photon-phonon interactions at the heart of OMOs working principle. As such, acoustic waves in the GHz have already been used to adress mechanical modes of optomechanical crystal for quantum applications \cite{forsch2020microwave} and strong coherent photon-phonon interaction \cite{balram2016coherent, li2015nanophotonic}.
Recently, injection locking of an optomechanical oscillator via acoustic waves has been also demonstrated, where the OMO has been locked to the frequency and phase of an external oscillator thanks to the acousto-optical interaction \cite{huang2018injection}.
\begin{figure}[h]
{\phantomsubcaption\label{fig1:a}}
{\phantomsubcaption\label{fig1:b}}
{\phantomsubcaption\label{fig1:c}}
\includegraphics[scale=0.31]{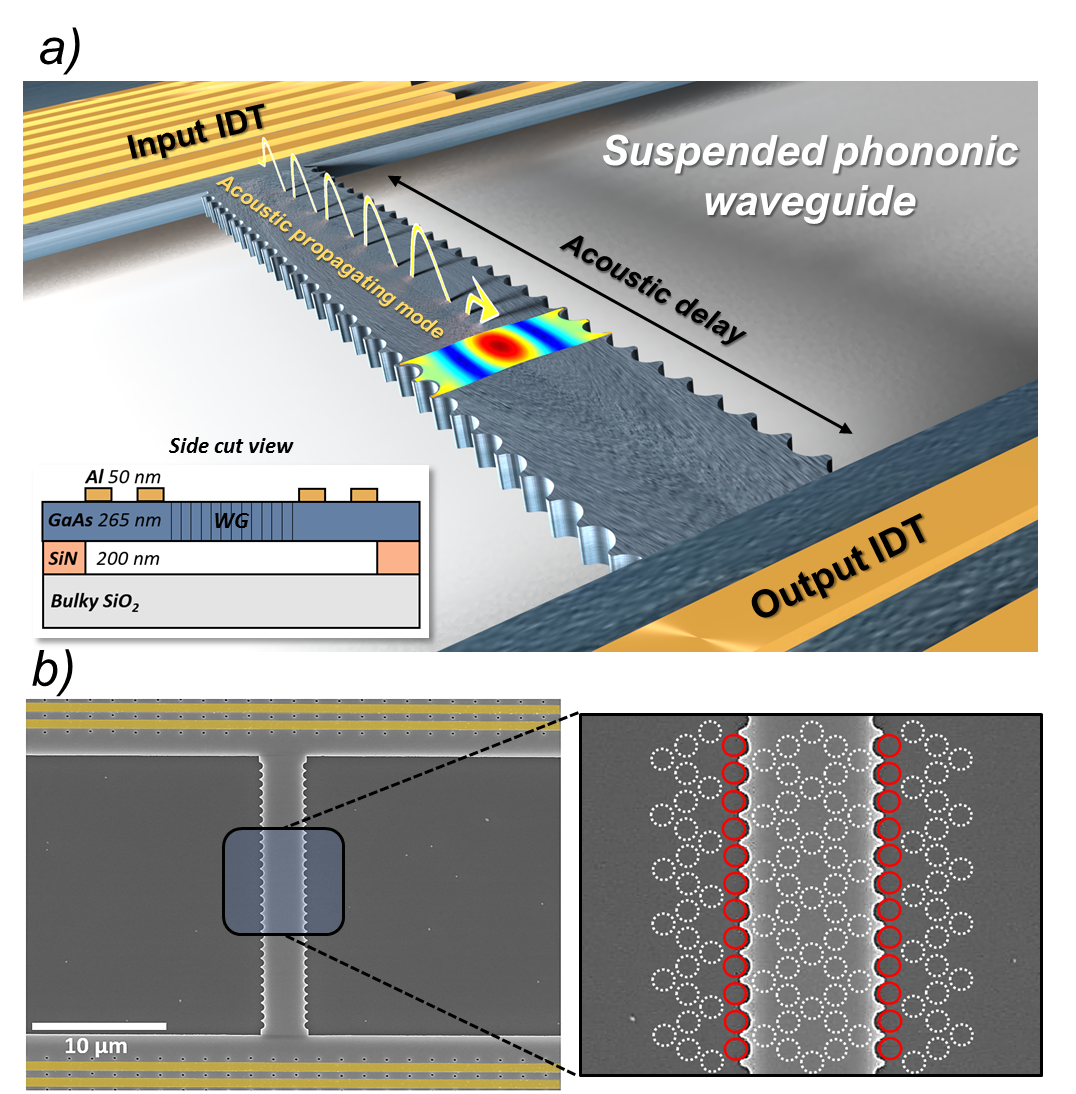}
\caption{a) Schematic of the system under study. Inset: side view cut of the structure  b) Top view SEM image of fabricated suspended phononic waveguide and IDTs (false color) in yellow. Inset zoom: Phononic waveguide design with Kagome crystal lattice implementation.}
\label{fig1}

\end{figure}
By taking advantage of the well-known low velocity of acoustical phonons in material as well as their immunity to enviromental disturbances, an on-chip acoustical feedback loop could be implemented. Even though, interest in guiding such waves with phononic waveguide in the wavelength scale structures increased \cite{dahmani2020piezoelectric, siddiqui2018lamb, shao2019phononic}, delay implementation in a controllable and compact manner is still a missing point. This prevented OMOs from reaching frequency stability and spectral purity comparable to OEOs.
Acoustic waves slowly propagating through phononic waveguides with tunable low velocity can play indeed an important role in the context of a fully-integrable on-chip OMO, allowing the implementation of the hundreads of ns delay needed for oscillations stabilization. Furthermore, a compact on-chip delay line could be also exploited for both classical and quantum information systems \cite{safavi2011proposal}, compact wireless communication devices \cite{lakin1995development} and phononic information processing \cite{olsson2008microfabricated}. \\
Here, we show the transmission of a slow acoustic propagating mode in the UHF regime through a suspended Gallium Arsenide (GaAs) phononic waveguide, bonded on silicon wafer. This waveguide acts at the same time as an acoustic filter.  
In order to investigate the dispersion engineering  for a specific mode within this acoustic waveguide, we use a pair of suspended interdigitated transducers (IDTs). Thus, by taking advantage of the inverse piezoelectric effect in GaAs, we evidence the filtering of the generated acoustical wave. Furthermore, with a time domain approach, we experimentally  obtain a group velocity below 1000 m/s thanks to group delay extrapolation of the acoustic transmitted signal. This open the way to a possible delay implementation  in a compact and fully-integrated microscale \textit{III-V} piezoelectric semiconductor optomechanical crystal.\\
       \\
       \\
The device under investigation is a 265nm thick GaAs waveguide placed between two IDTs made of Aluminium directly deposited on top of the GaAs slab (see \cref{fig1:a}). Silicon Nitride (SiN) is used as a sacrificial layer for the suspension of the structure (see inset \cref{fig1:a} for side cut view of the complete structure).
The system has been suspended mainly for two reasons: firstly, to confine the acoustical mode solely in the III-V material, and secondly, to avoid any leakage of the produced electric field towards the substrate.  
The fabrication process starts with the growth of the 265 nm GaAs slab by Metalorganic Chemical Vapor Deposition (MOCVD) on top of an Aluminum Gallium Arsenide (AlGaAs) layer on a GaAs substrate. 200 nm of SiN are then deposited by Plasma-Enhanced Chemical Vapor Deposition (PECVD) on the GaAs slab. 
At this point, an adhesive bonding with a thick SiO\textsubscript{2} layer (2 $\mu$m) on silicon wafer is performed using a polymerized divinylsiloxaane-benzocyclobutene (BCB) solution. Removal of both GaAs substrate and the etch-stop AlGaAs layer is then chemically obtained.
The fabrication of the electrodes constituting the IDTs is consequently carried out by electron beam lithography on a positive resist (PMMA) with following metal deposition and lift-off of 50 nm of Al. On a second lithography step the phononic waveguide is firstly drawn onto a PMMA resist. The pattern is then transfered by Reactive Ion Etching (RIE) technique to a SiN hard mask. This lattest layer is used to etch by Inductively Coupled Plasma (ICP) the GaAs thin slab with a gas mixture based on HBr. Subsequently, we use dry plasma based on SF\textsubscript{6} gas to underetch the SiN sacrificial layer in order to achieve the suspension. The phononic waveguide is underetched from its sides while the IDTs part is underetched through small holes opened in the GaAs layer, accurately placed between the IDTs fingers. A fully suspended structure consisting of a pair of IDTs and a phononic waveguide between them is finally achieved (see \cref{fig1:b}).

\begin{figure}[H]
\begin{center}
{\phantomsubcaption\label{fig2:a}}
{\phantomsubcaption\label{fig2:b}}
\includegraphics[scale=0.32]{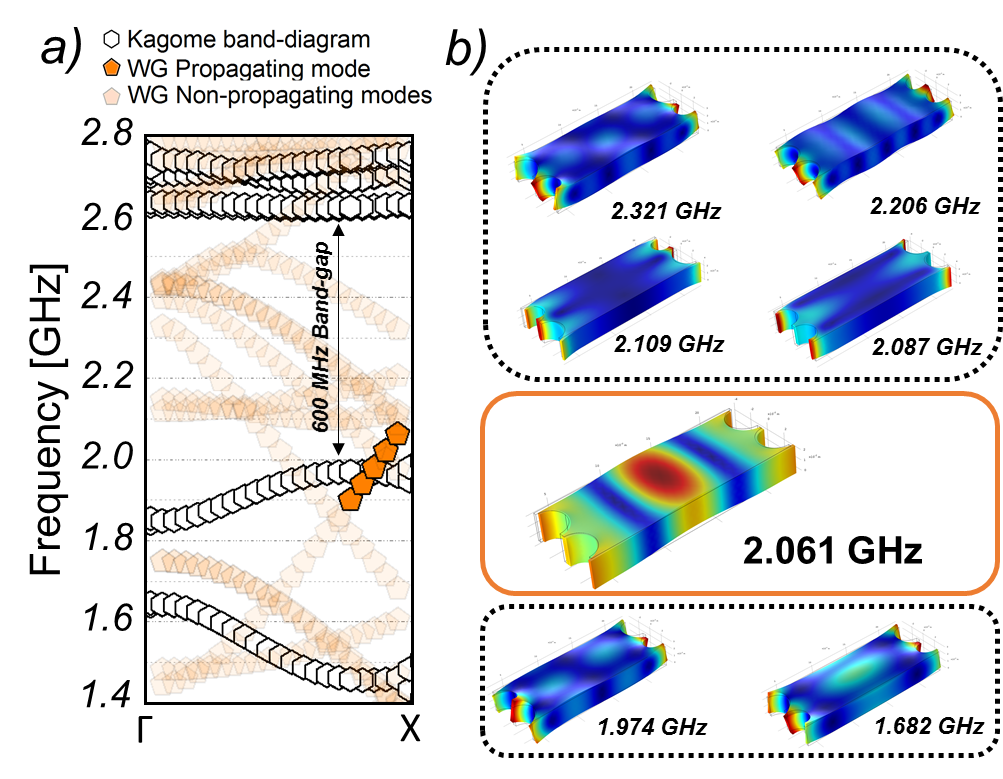}
\caption{a) FEM simulation of phononic band-gap for a Kagome crystal lattice with lattice constant \textit{a} = 370nm (white hexagons) and all existing modes along the direction of the waveguide created by removing 5 lines of holes (shaded and filled orange pentagons)  b) Total displacement of the mechanical propagating mode (orange frame) and non-propagating modes (black dashed frames).}
\label{thermoopt}
\end{center}
\end{figure}

Regarding the acoustic propagating mode, the out of plane confinement is ensured by the difference of acoustical properties between the GaAs slab and its surrounding. Lateral confinement is controlled and mastered by patterning a phononic crystal. Specifically, a Kagome crystal lattice design is exploited to engineer the phononic waveguide. In a 265 nm thick GaAs slab, a Kagome crystal lattice with lattice constant of 370 nm would lead to a phononic bandgap of 600 MHz in the range of 2 GHz (see white hexagons in \cref{fig2:a}) for a gap-to-midgap ratio of 26\%, wider with respect to other possible design implementations with 2D geometry such as triangular or hexagonal lattices \cite{2DPhNBG_mohammadi2007}. 
In our case, the waveguide has been created starting from perfect Kagome crystal lattice layout where 5 lines of holes have been removed, giving hence its width ($\simeq$ 1.7$\mu$m for a lattice constant \textit{a} = 370nm). We use, in addition, the two adjacent lines of holes on both sides to create indentations (see red circles in inset zoom \cref{fig1:b}).

Finite Element Method (FEM) simulation of all the values for the reduced wave vector \textit{k} propagating along the length of the waveguide has been carried out. 
For 5 missing lines of holes and a lattice constant of 370nm, every available mode in the waveguide is shown in \cref{fig2:a}, showing similar behavior with respect to waveguides in optical domain \cite{oser2019coherency}. In our case only one mode is propagating through the waveguide and is highlighted with orange pentagons in \cref{fig2:a}.
Total displacement of the well confined propagating mode at 2.061 GHz for \textit{k}=X is presented in the orange frame in \cref{fig2:b}. Here, it is furthermore shown that, in this chosen geometrical condition, all the other ones are indeed non-propagating modes (black dashed frames in \cref{fig2:b}). By taking into account that an odd number of holes must be removed in order to keep the waveguide symmetric, a wider waveguide (i.e. 7 or more missing holes) would bring degenerate modes, while a narrower waveguide (1 or 3 missing holes) would prevent a proper propagation and/or isolation of a unique acoustical propagating mode in the same range of frequency.
Beyond single frequency propagation, this designed waveguide also evidences slow mode features. By calculating the derivative of the reduced wave-vector \textit{k} values for the propagating mode the group velocity can be extracted as V\textsubscript{g}=$\frac{\partial \omega}{\partial k}$,  being $\omega$ the frequency. In this case, the simulated value for the group velocity related to the propagating mode turned out to be 606 m/s for $k$=X.

\begin{figure}[H]
\begin{center}
{\phantomsubcaption\label{fig3:a}}
{\phantomsubcaption\label{fig3:b}}
\includegraphics[scale=0.30]{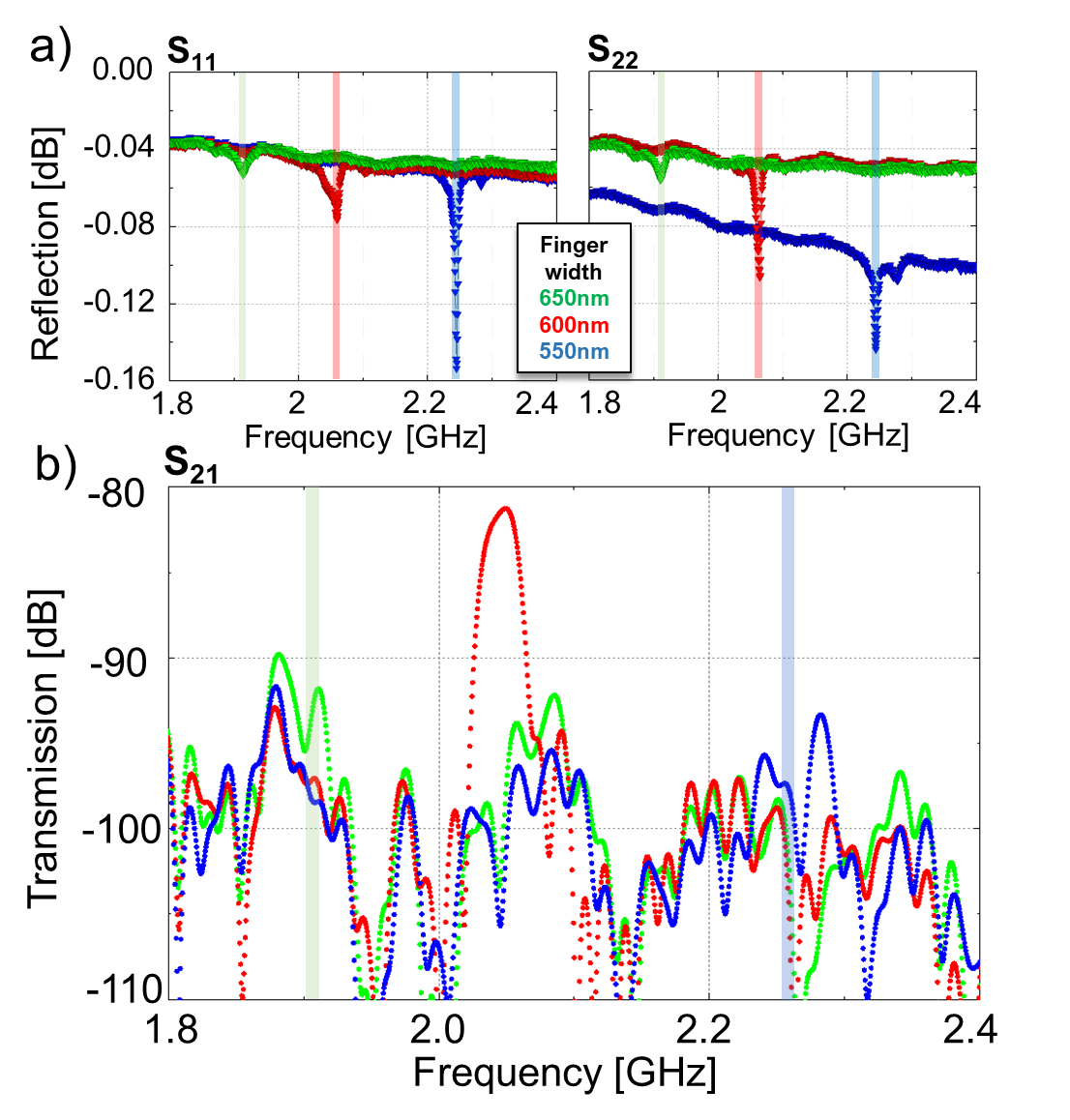}
\caption{a) Reflection coefficients for IDTs pairs with 650, 600 and 550 nm finger widths b) Transmission through a phononic engineered waveguide acoustically pumped at different frequencies. Shaded green, red and blu lines in the three graphs are for eye guiding.}
\end{center}
\end{figure}

On the acoustic actuation side, the design of the IDTs is performed in order to match the mechanical frequency of the propagating mode. Specifically, the resonance frequency of an IDT is determined by the elastic and piezoelectric properties of the chosen material, its thickness and the chosen finger width for the electrodes pair. In order to test acoustic transmission at several frequencies around 2 GHz, we have fabricated structures with 650, 600 and 550 nm as finger width. A 2-ports VNA has been used to extract S-parameters by applying a power of 0 dBm to each transducer. The reflection coefficients S\textsubscript{11} and S\textsubscript{22} of the tested set of IDTs are shown in \cref{fig3:a}. Here, resonances at 1.91, 2.06 and 2.245 GHz are found for 650, 600 and 550 nm finger width respectively for both S\textsubscript{11} and S\textsubscript{22}, showing that the transducers are working at exactly the same frequency on both sides, i.e. transmission signals would be expected to be properly transduced for each case.
The transmitted signals between two IDTs through a 20 $\mu$m long phononic waveguide are shown in \cref{fig3:b}. Here, the S\textsubscript{21} traces have been gated within a 13-75 ns time interval. In fact, time-gating technique is needed to reveal the transmission, getting rid of the electromagnetic radiation (EMR) between the electrical probes in our experimental setup. We observe in \cref{fig3:b} transmission only for the structure with 600 nm finger width, i.e. when the provided acoustical pumping frequency (=2.06 GHz) is matching the frequency of the designed propagating mode (=2.061 GHz) of the waveguide. Hence this demonstrates that only one propagating mode exists in the range of the three different tested transducers and that it can be indeed properly acoustically pumped. In other words, the phononic waveguide is acting as an acoustic filter, filtering out the transmissions for the IDTs working at 1.91 and 2.245 GHz, allowing only the transmission of a signal matching the frequency of the designed acoustic propagating mode.

\begin{figure}[H]
\begin{center}
{\phantomsubcaption\label{fig4:a}}
{\phantomsubcaption\label{fig4:b}}
\includegraphics[scale=0.35]{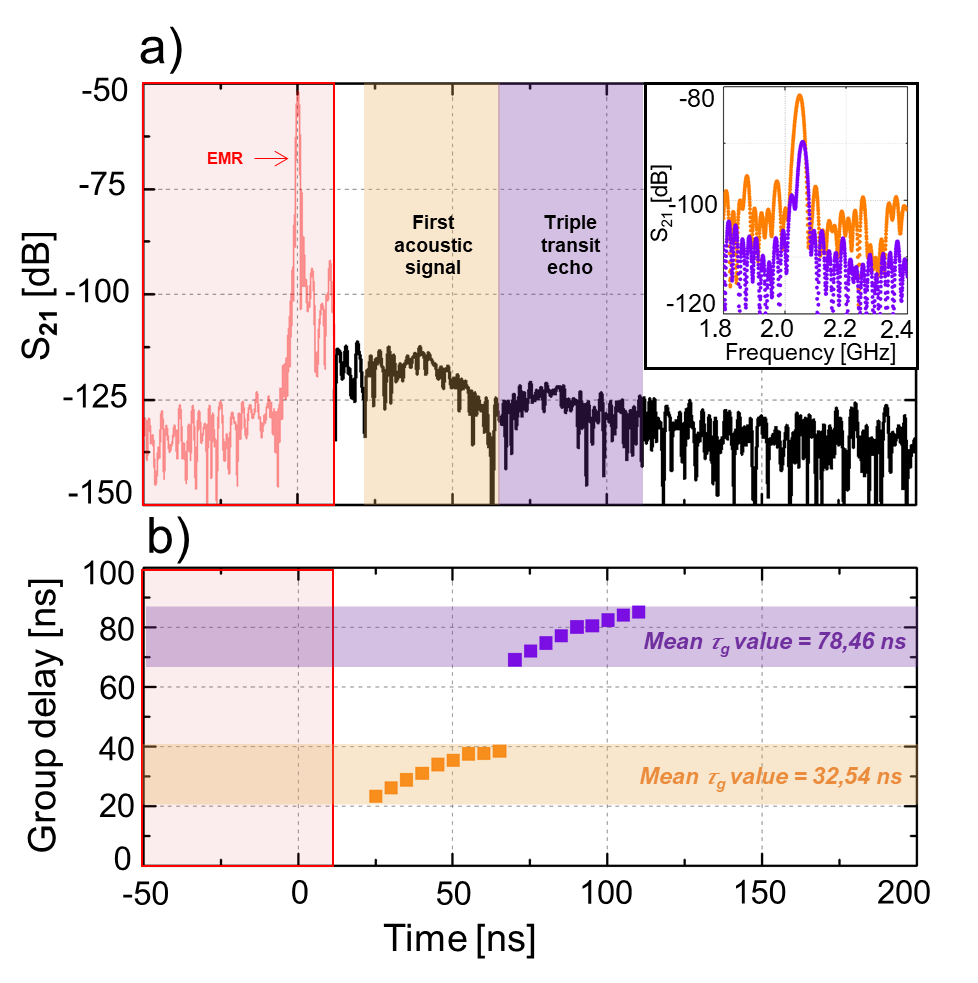}
\caption{a) S\textsubscript{21} in time-domain. The shaded red part contains the EMR peak in proximity of t = 0 ns. The shaded orange time window contains the first acoustic signal, the shaded violet time window contains the TTE. Inset: Orange and violet time-windows transformed back to frequency domain revealing the related transmission peaks. b) Group delay at peaks value. Time-gate window starts at 20 ns for the first acoustic signal and at 65 ns for the TTE. Values are sampled every 5 ns.}
\end{center}
\end{figure}

In order to experimentally estimate the velocity, we study the time domain behaviour of S\textsubscript{21}. \Cref{fig4:a} shows the time domain impulse response of the system under study. 
Time-gating reveals the transmission signals filtering out the EMR that manifests in frequency domain as a flat wide-band signal severely reducing the out-of-band insertion loss \cite{clement2004saw}. In time domain, the EMR shows itself as an intense narrow peak after a very short delay (electromagnetic signal through the substrate or air propagates much faster than acoustic wave) as visible in \cref{fig4:a} at t = 0 ns. Two acoustic signals can be discriminated in the time domain trace. The first acoustic signal arriving to the output IDT directly from the input IDT and the triple-transit echo (TTE), which has traversed three times the path between input and output IDTs. We specifically time-gate the two signals with two time-gate windows sufficiently wide to contain the acoustic components related to both of them (shaded orange and violet frames in \cref{fig4:a}). By Fourier-transforming back to frequency domain  the two time domain segments we filter out the EMR and the two signals can be revealed (inset \cref{fig4:a}).
We can study at this point the group delay in order to experimentally estimate the group velocity of the acoustic propagating mode. Specifically, the group delay can be defined as $\uptau\textsubscript{g}$ = -$\frac{\partial \phi\textsubscript{rad}}{\partial \omega}$ where $\phi$\textsubscript{rad} is the phase of the S\textsubscript{21} transmission response.

By fixing the starting point of the time-gating window at 20 ns and 65 ns for the two acoustical signals and gradually increasing by 5 ns the width of these windows, we extract the group delay for both signals (\cref{fig4:b}).
The mean $\uptau\textsubscript{g}$ value for the first acoustic signal and the triple transit echo is 32.54 ns and 78.46 ns respectively. Being 24 $\mu$m the distance between the two IDTs (20 $\mu$m of waveguide plus 2 $\mu$m distance between waveguide and IDT on each side) the calculated group velocity is then 737 $\pm$ 123 m/s for the first acoustic signal and 917 $\pm$ 153 m/s for the TTE that has traversed a 72$\mu$m path. Here the uncertainty value belongs to the velocity uncertainty in the 2 $\mu$m per side ($\simeq$ 17$\%$ of the total path) that have not been phononically engineered. These experimental results are in the same order of magnitude of the simulated value of 606 m/s for the group velocity discussed before. It follows that this combination of IDTs and phononic waveguide can be used to introduce a delay of 1.36 $\pm$ 0.23 ns/$\mu$m  in a fully-integrable manner.
\\
In conclusion, we have demonstrated propagation and filtering of slow acoustical waves in the UHF regime through a suspended and patterned membrane embedding a phononic waveguide made of GaAs on an insulator wafer. Simulation, fabrication and measurements have been shown, with particular emphasis to the specific phononically engineered propagating mode at 2.06 GHz. Thus, we evidence that a fully integrated on-chip delay in the order of few ns/$\mu$m is at reach. Being by essence insensitive to environmental disturbances, this can set the ground to drastically enhance performances of compact engineered nano-structures for information processing where a delay is needed such as integrated low noise optomechanical oscillators. 
\\
\\
\\

G.M. acknowledges support from the European Union's H2020 research and innovation programme under the Marie Skłodowska-Curie grant agreement No. 722923 (OMT). This work is supported by the French RENATECH network,  the European Union’s Horizon 2020 research innovation program under grant agreement No 732894 (FET Proactive HOT) and the Agence Nationale de la Recherche as part of the “ASTRID” program (CRONOS, ANR-19-ASTR-00-22-01)
\\
\\
The data that support the findings of this study are openly available at 10.5281/zenodo.3888069.

\bibliography{Acoustic_bib}
\end{document}